\documentclass[twocolumn,preprintnumbers,amsmath,amssymb]{revtex4}

\usepackage{graphicx}
\usepackage{dcolumn}
\usepackage{bm}

\begin{document}

\preprint{NStaley}

\title{Manipulating superconducting fluctuations by the Little-Parks-de Gennes effect in ultrasmall Al loops}

\author{Neal E. Staley}
\author{Ying 
Liu}
\email{liu@phys.psu.edu}
\affiliation{%
Department of Physics and the Institute of Materials Research, 
The Pennsylvania State University, University Park, PA 16802, 
U.S.A.
}%
\date{\today}

\begin{abstract}
The destruction of superconducting phase coherence by quantum fluctuations and the control of these fluctuations have been a problem of long-standing interest, with recent impetus provided by its relevance to the pursuit of very high temperature superconductivity. Building on the work of Little and Parks, de Gennes predicted more than three decades ago that superconductivity could be destroyed near half-integer-flux quanta in ultrasmall loops with a side branch, resulting in a destructive regime. We report the experimental observation of this Little-Parks-de Gennes effect in Al loops prepared by advanced e-beam lithography. We show that the effect can be used to restore the lost phase coherence through side branches. 
\end{abstract}

\maketitle

The existence of a many-body state composed of fluctuating, phase-incoherent Cooper pairs has been a problem at the forefront of superconductivity research for decades. Such a state, which may emerge when the global superconducting phase coherence is destroyed by strong disorder, Coulomb repulsive interaction, or magnetic field \cite{LarkinFluctTheory}, may have been realized in the pseudogap phase of high critical temperature (T$_c$) superconductors found at temperatures well above T$_c$ \cite{XuNernst2000}. If the fluctuations in such a state could be suppressed and global phase coherence built through engineering means, very high temperature superconductivity would be obtained. A similar state with a different physical origin, to be discussed in detail below, is expected in doubly connected ultrasmall superconductors \cite{deGennesLoop, StraleyLoops, LiuScience01}. Recent theoretical studies have produced some detailed predictions on the nature of the thermal and quantum superconducting fluctuations in this state, and how the fluctuations may be controlled \cite{ShahFluctPRB07, SchwietePRL09, SchwietePRB10}. Novel phenomena, such as $hc/e$ as opposed to the $hc/2e$ Little-Parks oscillations as the size of the loop is reduced \cite{WeiGoldbartRing, VakaryukRing} and the occurrence of superconductivity in the smallest doubly connected samples, the thinnest carbon nanotubes \cite{TangSuperNanotube}, may be expected. New experimental techniques capable of probing and manipulating these fluctuations were developed \cite{KoshnickSquidFluctScience07, StewartVallesSIHoneycomb}. Therefore, similar to singly connected mesoscopic superconductors in which some spectacular physical phenomena were found \cite{GeimNature97, ChibotaruNatureAntivortex, HopkinsBezryadin}, ultrasmall doubly connected ultrasmall superconductors are likely to grow into a fertile testing ground for fundamental studies of superconductivity, including the pursuit of very high temperature superconductivity. 

\begin{figure}
\includegraphics[ scale 
=1.0]{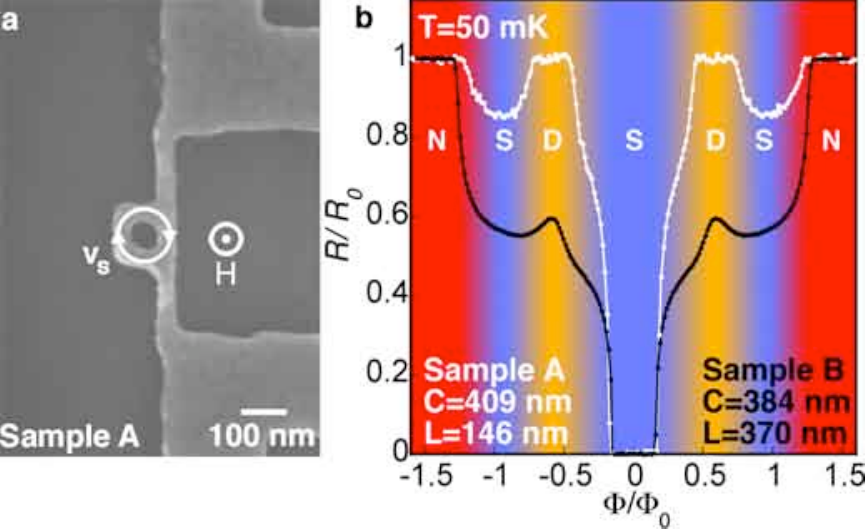}
\caption{(Color Online) a) Scanning Electron Microscope (SEM) image of a representative sample. The superfluid velocity, v$_s$, induced by the magnetic field, H, is shown schematically; b) Resistance (R) vs. $\Phi/\Phi_0$ ($\Phi_0$ = $hc/2e$) at 50 mK for two samples with their loop circumference $C$ and lead length $L$ indicated. The superconducting (S), destructive regime (D), and normal (N) states are shown. Zero resistance is not expected at $\Phi_0$ (corresponding to a field of ~2000 G for these samples) because the field is close to the critical field and the wide parts of the leads are normal.  }
\end{figure}

\begin{figure}
\includegraphics[ scale 
=1.0]{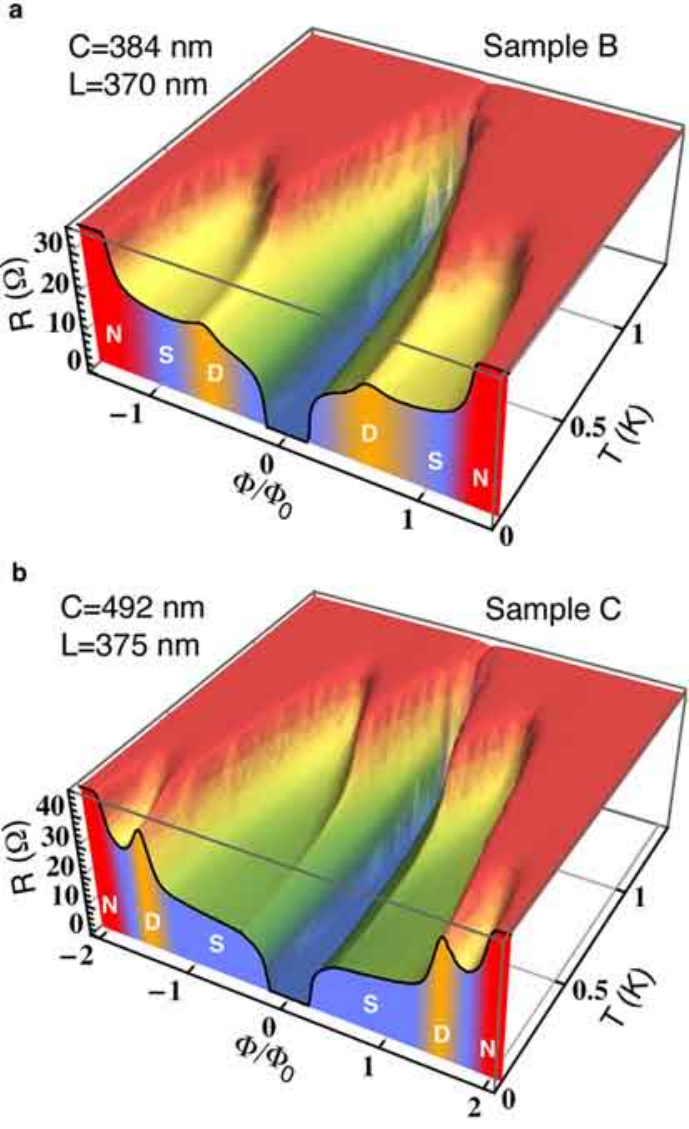}
\caption{(Color Online) Resistance (R) $vs.$ magnetic flux ($\Phi/\Phi_0$) and temperature (T) for a sample with a circumference $C$ = 384 nm and lead length $L$ = 370 nm (a) showing the destructive regime near $\Phi_0$/2 and a sample of $C$ = 492 nm and $L$ = 375 nm (b) showing a destructive near 3$\Phi_0$/2, but not near $\Phi_0$/2. Each plot was constructed from R vs. $\Phi/\Phi_0$ traces taken at 100 mK intervals from 50 mK to 1.4 K.  }
\end{figure}

\begin{figure}
\includegraphics[ scale 
=1.0]{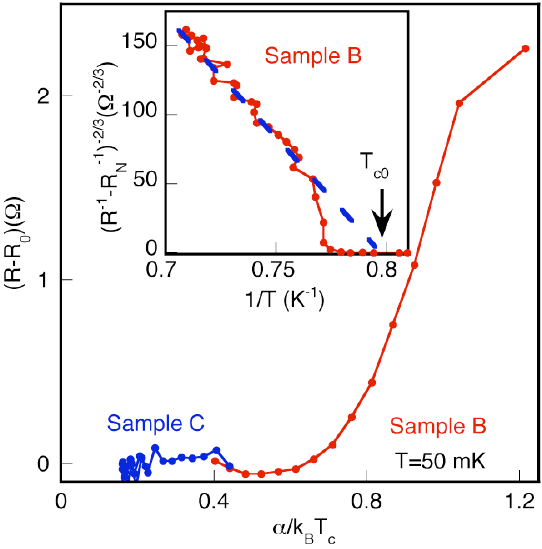}
\caption{(Color Online) Resistance change as a function of the dimensionless depairing energy $\alpha/k_BT$ at 50 mK between (0.5 - 1) $\Phi/\Phi_0$ for two loops with different $C$ but similar $L$ values. The inset shows change in conductance $vs.$ 1/T at zero field for the sample above T$_{c0}$. The dashed line indicates fit to the linear behavior expected from Aslamasov-Larkin theory for fluctuation enhanced conductance in one dimension\cite{Tinkham}, yielding T$_{c0}$= 1.25 K consistent with that obtained directly.}
\end{figure}

\begin{figure}
\includegraphics[ scale 
=1.0]{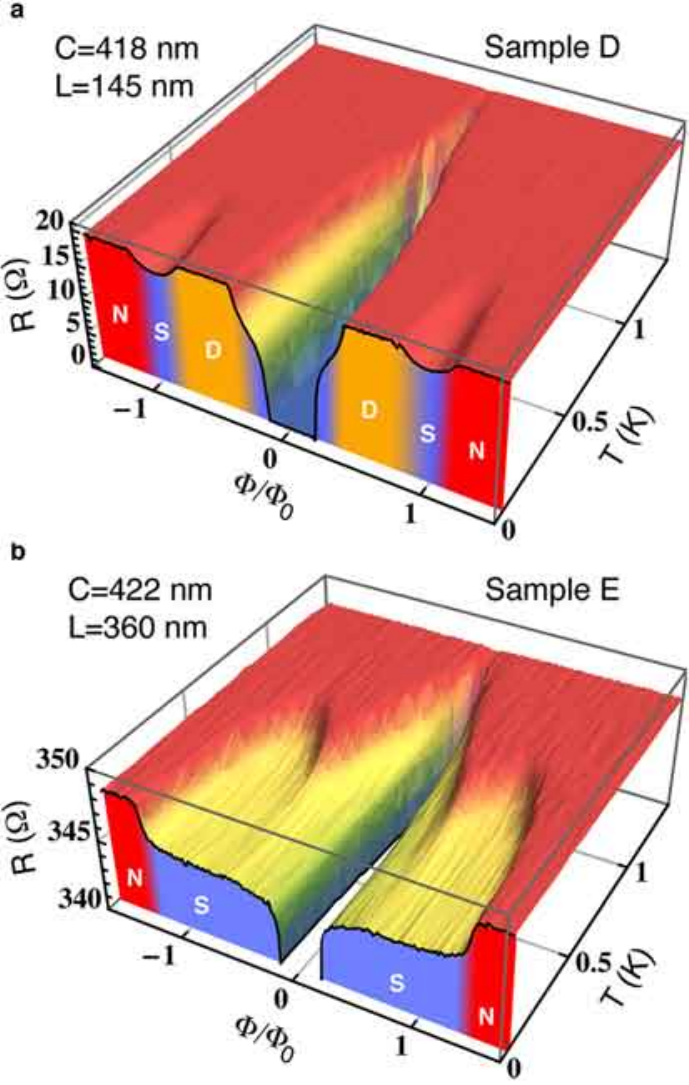}
\caption{(Color Online) R $vs.$ $\Phi/\Phi_0$ and T of a sample with $L$ = 145 nm and $C$ = 418 nm (a) showing a robust destructive regime with full normal state resistance over a wide flux range (0.4 - 0.7 $\Phi_0$ at 50 mK) and a sample of $L$ = 360 nm and $C$ = 422 nm (b) showing no destructive regime near $\Phi_0$/2. Each plot was constructed from R $vs.$ $\Phi/\Phi_0$ traces taken at 100 mK intervals from 50 mK to 1.4 K. Even though Sample E is a 3-terminal sample due to a damaged lead, it was chosen for comparison because Samples D and E were prepared under identical conditions and positioned next to one another on the same chip, ensuring that their sample parameters are as close as possible.}
\end{figure}

The fluxiod quantization in a thin, doubly connected superconductor requires that an applied flux ($\Phi$) threading the superconductor, produce a superfluid velocity $v_s\sim (1/C)(n-\Phi/\Phi_0)$, where $C$ is the circumference of the loop, $n$ is an integer that minimizes $v_s$, and $\Phi_0$(= $hc/2e$) is the flux quantum. The periodic modulation of $v_s$, which reaches a maximum at half-integer-flux quanta, leads to the well-known Little-Parks oscillations in the superconducting transition temperature (T$_c$) \cite{LittleParks, TinkhamLittleParks}. Based on a Ginzburg-Landau theory, de Gennes made a prediction in 1981 \cite{deGennesLoop} that the T$_c$ oscillation amplitude could become so large for ultrasmall samples that superconductivity itself is destroyed (T$_c$ = 0) near half-integer-flux quanta due to the growth of maximal $v_s$ and the associated kinetic energy as $C$ shrinks. The resulting non-superconducting state is referred to as the destructive regime\cite{deGennesLoop, StraleyLoops, LiuScience01}.  Furthermore, de Gennes predicted that the addition of a ``dangling'' side branch, which does not change $v_s$ and naively should not affect superconductivity in the loop, could stabilize the phase coherence in the entire sample and suppress the destructive regime\cite{deGennesLoop, StraleyLoops}. The presence of the destructive regime and its suppression by the side branch in ultrasmall superconducting loops are referred to here as the Little-Parks-de Gennes (LPdG) effect. 

In order to observe the LPdG effect the circumference, $C$, of the superconducting loop must be smaller than $\pi\xi$(0), where $\xi$(0) is the zero temperature superconducting coherence length with typical value around 100 nm for superconducting microstructures. Preparation of such an ultrasmall loop is a significant challenge. Previous studies of superconducting loop structures almost all featured a minimal loop size around 1 $\mu$m, far too large to observe the LPdG effect \cite{SanthanamRB1989Loops}. While the use of doubly connected ultrathin cylinders \cite{LiuScience01, HaohuaPRL05, HaohuaThesis, SternfeldPRL2011} led to the successful demonstration of the destructive regime, the full LPdG effect, especially the role played by the side branch has not been explored experimentally. Our samples were prepared by advanced electron beam lithography, yielding a line width of 40 nm with 31 nm thick Al.  The electrical transport measurements were carried out at d.c. technique in an RF filtered dilution refrigerator with a base temperature of 20 mK. Two electrical leads connected to the loop (Fig. 1a) conveniently function as the side branches in de Gennes' theory, whose effective length, $L$, is essentially that of the narrow part of the leads. Our devices showed a low level of disorder, with a zero-field transition temperature T$_{c0}$ = 1.28 K and $\xi$(0) = (100 $\pm$ 15) nm\cite{Note}.

The destructive regime was observed in our ultrasmall loop structures.  Shown in Fig. 1b are resistance $vs.$ magnetic flux traces taken at 50 mK for two loops with circumferences $C$ = 409 and 384 nm and side branches of a length $L$ = 146 and 370 nm, respectively. At zero field the entire device is superconducting, and with increasing field the wide parts of the leads are driven normal first, at a field of 350 G. As the applied field increased further, with the corresponding flux approaching to $\Phi_0$/2, the loop is driven into the normal state, marking the emergence of a superconductor-normal metal quantum phase transition (QPT). While Sample A reaches the full normal state resistance between 0.45 $\Phi_0$ and 0.7 $\Phi_0$, Sample B features a resistance peak at $\Phi_0$/2, demonstrating in both cases the presence of a destructive regime and QPT. 

The flux tuned QPT in these ultrasmall Al loops can be manipulated by varying the loop circumference, $C$.  Shown in Fig. 2 is data obtained from two samples of different $C$ values but a similar lead length, $L$ $\approx$ 370 nm.  In Fig. 2a, a $C$ = 384 nm sample was found to show a distinctive resistance peak near $\Phi_0$/2, indicating the destruction of superconductivity.  For the sample with $C$ = 492 nm, however, superconductivity remains robust at $\Phi_0$/2 as shown in Fig. 2b.

The attainment of the destructive regime made it possible to examine recent predictions regarding quantum fluctuations in the destructive regime based on pair breaking theories \cite{ShahFluctPRB07, SchwietePRL09, SchwietePRB10}, which provided a framework for understanding fluctuation enhanced electrical conductivities \cite{ShahFluctPRB07}, or the magnetic responses from the fluctuating persistent currents \cite{SchwietePRL09, SchwietePRB10} in the destructive regime. Within this theory, for loop structures at a given magnetic flux, the energies for a pair of time-reversal single particle states differ by $\alpha$ \cite{GroffParks}, which quantifies the energy difference for two time-reversal single particle states. A reduced T$_c$ is obtained by solving the equation
\begin{equation}
\ln\left(\frac{T_c(\alpha)}{T_{c0}}\right)=\psi\left(\frac{1}{2}\right)-\psi\left(\frac{1}{2}+\frac{\alpha}{2\pi k_BT_c(\alpha )}\right)
\end{equation}
where $\psi$ is the digamma function\cite{ShahFluctPRB07, SchwietePRL09, SchwietePRB10}.  Including second order effects related to a nonzero wire width\cite{SchwietePRB10}, we have 
\begin{equation}
\begin{split}
\alpha=&\frac{\hbar D}{2 R^2}\biggl(\left(\frac{\Phi}{\Phi_0}-n\right)^2\\
&+\frac{w^2}{4R^2}\left[\left(\frac{\Phi}{\Phi_0}\right)^2+n^2\left(\frac{1}{3}+\frac{w^2}{20R^2}\right)\biggr]\right)
\end{split}
\end{equation}
where $D=\frac{8k_BT_{c0}\xi^2(0)}{\pi\hbar}$ is the diffusion coefficient for disordered superconductors, and $w$ is the line width. In this theory, the quantum critical point is given by T$_c = 0$ of the ring, yielding $\alpha_c=0.889k_BT_{c0}$. Similarly in mean field Ginzburg-Landau theory, \cite{GroffParks}
\begin{equation}
\begin{split}
\frac{\Delta T_c}{T_{c0}}=&\frac{4\pi^2\xi^2(0)}{C^2}\biggl[\left(\frac{\Phi}{\Phi_0}-n\right)^2+\left(\frac{\Phi}{\Phi_0}\right)^2\left(\frac{\pi w}{C}\right)^2\\
&+n^2\left(\frac{\pi w}{C}\right)^2\left(\frac{1}{3} +\frac{1}{5}\left(\frac{\pi w}{C}\right)^2\right)\biggr]
\end{split}
\end{equation}
Taking $\Delta T_c$ to T$_{c0}$, we arrived the condition for T$_c=0$.  In the case of the destructive regime is not reached, the value of T$_c$ can be predicted using both theories.  A comparison between these predictions and our experimental results is shown in Table 1 for Sample C.  Both theories have the correct trend, especially the presence of the destructive regime at 3$\Phi_0$/2 but not at $\Phi_0$/2 in Sample C but neither predicts the experimentally observed destructive regime in this sample as shown in Table 1.  It should be emphasized that in both cases the effects of the addition of two side branches were not considered.

\begin{table}
\caption{\label{tab:table1}{Comparison of pair breaking theory and mean-field Ginzburg-Landau theory predictions to the experimental results for Sample C.  }}
\begin{ruledtabular}
 \begin{tabular}{l | c c c c}
Flux &$\alpha/k_BT_{c0}$&T$_{c}(\alpha) (mK)$&T$_{c} (GL) (mK)$&T$_{c }(Meas) (mK)$\\
\hline
$\Phi/2\Phi_0$&0.53&700 &750 &320 \\
$3\Phi/2\Phi_0$&0.87&100 &400 &$<$50 \\
\end{tabular}
\end{ruledtabular}
\end{table}

The reduced sample resistance seen in the destructive regime (Fig. 1b) clearly comes from superconducting fluctuations, a situation different from that in the previous cylinder work where similar finding could in principle be due to sample specific causes \cite{HaohuaThesis, VafekLoopInhomogeneity}. Here the thermal superconducting fluctuations expected above the mean-field transition, T$_{c0}$, were explicitly demonstrated (Inset of Fig. 3).  However, the analysis of quantum fluctuations, predicted to depend on depairing energy, $\alpha$, which increases with $C^{-2}$\cite{ShahFluctPRB07}, needs to be handled carefully. The main issue is determining when quantum fluctuations start to dominate, which is not clear even theoretically \cite{ShahFluctPRB07, SchwietePRL09, SchwietePRB10}. However, given the large values of $\alpha$ because of the small $C$ value, at 50 mK, which corresponding to T/T$_{c0}$ = 0.039, quantum fluctuations should be at least substantial in our ultrasmall samples. This is consistent with the behavior observed in samples with different circumference but a fixed lead length in which the reduced resistance seems to be a function of depairing energy $\alpha$, as shown in Fig. 3 for Sample B and Sample C. It is interesting to also note that, when the finite width correction is considered, an additional broadening in $\alpha$ is apparent in Fig. 3, as expected \cite{SchwietePRB10}. 

A central feature of the LPdG effect, that the destructive regime is strongly affected by the presence of a side branch on the loop \cite{deGennesLoop, StraleyLoops}, was readily tested in the current experiment. In Fig. 4 we show results obtained from two loop structures with similar $C$ but different $L$ values. While the sample with short side branches showed a robust destructive regime with full normal state resistance over a wide range of flux, the other with long side branches was found to be fully superconducting near $\Phi_0$/2 at the lowest temperatures. Therefore a sufficiently long side branch does suppress the destructive regime, as predicted \cite{deGennesLoop, StraleyLoops}. Evidently the side branch functions as a reservoir of condensation energy, which balances out the kinetic energy rise in the loop and helps restore the global phase coherence by converting fluctuating pairs into a superconducting condensate.   

The QPT seen in our ultrasmall Al loops is different from the two-dimensional superconductor-insulator transitions (SITs) studied previously \cite{GoldmanSITBook}. Here the sample is of a finite size and therefore long-range phase coherence is not a defining feature of the state on either side of the QPT. Such a QPT has been encountered previously in finite-size systems \cite{PenttilaSonin}. Another fundamental question concerns the nature of the quantum superconducting fluctuations in the destructive regime.  In almost all known cases, the origin of the quantum superconducting fluctuations can be traced to the uncertainty relation \cite{Tinkham}, $\Delta N\Delta\phi>1$, which attributes the fluctuations in the phase of the superconducting order parameter ($\phi$) to reduced fluctuations in the number of Cooper pairs ($N$). In our samples, no mechanism to reduce $\Delta N$ is present even though $\Delta\phi$ can clearly be large. A different mechanism for inducing quantum fluctuations must then be at work. It is an intriguing possibility that the observed quantum fluctuations are mainly amplitude rather than the phase fluctuations. Amplitude fluctuations were discussed previously \cite{LarkinFluctTheory} in the context of two-dimensional SITs \cite{LarkinFluctTheory, GoldmanSITBook} and in high-T$_c$ superconductors \cite{KivelsonReviewStripes}. In these previous cases, fluctuations of phase and amplitude tend to be intermixed and difficult to analyze. Ultrasmall superconducting structures may then provide us with a simple model system in which quantum fluctuations may be accurately controlled by device design and applied flux. The restoration of lost phase coherence among fluctuating pairs by the addition of a side branch suggests that global phase coherence could be engineered, a possible pathway to very high temperature superconductivity. 

We would like to acknowledge useful discussions with J. Jain, A.M. Goldman, C-C. Tsuei, Y. A. Ying and C. P. Puls. We would also like to thank G. Harkay for translating Ref. \cite{deGennesLoop}.  This work is supported by NSF under Grant DMR 0908700 and Penn State MRI Nanofabrication Lab under NSF Cooperative Agreement 0335765, NNIN with Cornell University.
\bibliography{Bib.bib}

\begin{thebibliography}{29}
\bibitem[{\citenamefont{Staley}()}]{Note}
\bibinfo{author}{\bibfnamefont{The values of $\xi$(0) are estimated using the measured geometry of each sample and from the upper critical field of the devices.  In addition $\xi$(0) estimations from the measured resistivity were performed and yielded similar results.}}
\end{thebibliography}

\end{document}